\documentstyle[11pt,newpasp,twoside,epsf]{article}
\def\edcomment#1{\iffalse\marginpar{\raggedright\sl#1\/}\else\relax\fi}
\reversemarginpar

\begin{document}
\title{Type Ia Supernovae and the Value of the Hubble Constant}
\author{Brad K. Gibson, Chris B. Brook}
\affil{Centre for Astrophysics \& Supercomputing, Swinburne University,
Mail \#31, P.O. Box 218, Hawthorn, Victoria, 3122  Australia}

\begin{abstract}
The methodology involved in deriving the Hubble Constant via the 
calibration of the corrected peak luminosities of Type~Ia supernovae (SNe)
is reviewed.  We first present a re-analysis of the Cal\'an-Tololo (C-T) 
and Center for Astrophysics (CfA) Type~Ia SN surveys.  
Bivariate linear least squares and quadratic boot-strapped fits in peak 
apparent magnitude and light curve shape are employed to correct
this heterogeneous sample of peak apparent magnitudes, resulting in
an homogeneous (and excellent) secondary distance indicator: the so-called
corrected peak luminosity.
We next provide an empirical calibration for this corrected luminosity,
using Cepheid-based distances for seven nearby spiral galaxies
host to Type~Ia SNe.  Included in this sample is the
spectroscopically peculiar SN~1991T (in NGC~4527), whose corrected
peak luminosity is shown to be indistinguishable from that of so-called
``normal'' SNe.  A robust value of the Hubble Constant is derived and
shown to be H$_0$=73$\pm$2(r)$\pm$7(s)\,km\,s$^{-1}$\,Mpc$^{-1}$.
\end{abstract}

\section{Introduction}

The maximum brightness of a Type~Ia supernova (SN) is \it not \rm
a standard candle.  As illustrated graphically by Hamuy et~al. (1996) and
Riess et~al. (1996), there is an order of magnitude span 
seen in the luminosities at the peak of the SNe light curves in their
respective
samples (depending upon the bandpass in question).  On the other hand, an 
outstanding standard candle \it can \rm be constructed by applying
a light curve shape-dependent correction factor to the observed peak
magnitude (e.g., Phillips et~al. 1999; Jha et~al. 1999; Suntzeff
et~al. 1999; Gibson et~al. 2000; Freedman et~al. 2001; Gibson \&
Stetson 2001).  
This corrected peak luminosity succeeds as a standard candle where others 
fail because (i) the extreme luminosity of Type~Ia
SNe allows one to probe substantially further into the unperturbed Hubble
flow (in comparison with competing secondary indicators such as the
Tully-Fisher relation or surface brightness fluctuations), and (ii) the
intrinsic scatter is smaller than that of other indicators.  

Calibrating this secondary distance indicator can be approached from
a purely theoretical tack (e.g., H\"oflich \& Khokhlov 1996), but 
the numerous model uncertainties still lead the conservative to 
favour the empirical calibration.
The \it Sandage/Tammann/Saha 
Type~Ia SN HST Calibration Project \rm (Saha et~al. 1999)
was designed to provide just such a calibration, by
determining the Cepheid distances to eight nearby galaxies (IC~4182,
NGC~5253, 4536, 4496A, 4639, 3627, 3982, 4527) host to Type~Ia SNe.  This
original program has been supplemented with two additional calibrators
situated within NGC~3368 (Tanvir et~al. 1999) and NGC~4414 (Turner et~al.
1998).  Table~1 of Gibson et~al. (2000) provides a complete list of the
galaxies and SNe covered by all these calibration programs, including a
subjective quality ranking.

In what follows, we review the steps involved in calibrating the
Type~Ia SNe extragalactic distance scale.  We first provide a consistent
set of linear and quadratic fits to the peak magnitude-light curve
shape data from the Cal\'an-Tololo (Hamuy et~al. 1996)
and Centre for Astrophysics (Riess et~al. 1998,1999) SNe samples.  These
fits are then shown to provide a robust measure of the Hubble Constant when
the relations are calibrated using the seven best (currently available)
nearby Type~Ia SNe with Cepheid-based host galaxy distance 
determinations.

\section{Constructing a Standard Candle}

\it If \rm the peak luminosity of Type~Ia SNe were a standard candle, one would
expect the dispersion of the linear Hubble Diagram (magnitude versus
redshift) to be limited only by photometric precision, providing both
a precise \it and \rm accurate value for the Hubble Constant H$_0$ (once the
zero point was set by calibrating the peak luminosity via nearby
SNe of \it known \rm distance).  Combining the 50 Type~Ia 
SNe\footnote{SN~1996ab is not considered in what follows as its redshift is
in excess of the regime over which the $k$-corrections employed by
Hamuy et~al. (1996) are applicable.} in the Cal\'an-Tololo (C-T:
Hamuy et~al. 1996) and Center for Astrophysics (CfA: Riess et~al. 1998,1999)
samples, Freedman et~al. (2001; Fig~4) show that the raw B-, V-, and I-band
Hubble Diagrams are anything but indicative of a standard candle (e.g., the
dispersion in the B-band corresponds to a dispersion in distance (and
H$_0$) in excess of 40\%).  Even upon applying corrections for
foreground and host galaxy reddening, and culling the C-T$+$CfA samples
for low-redshift interlopers (which are more susceptible to peculiar
velocity corrections) and highly-reddened SNe, the resulting dispersion and
structure seen in the Hubble Diagrams (middle column of Fig~4 of
Freedman et~al. 2001) still indicate that an adequate standard candle is
not at hand.  To provide a useful constraint on H$_0$,
it is clear that these raw Hubble Diagrams need modification.

As has been well-documented over the past five years (e.g., 
Hamuy et~al. 1996; Riess et~al. 1996), 
the necessary modification
is provided by applying a mild correction to the observed peak magnitudes
of the Type~Ia SNe, the size of the correction being a function of
the shape of the light curve.  In its simplest form (Hamuy et~al. 1996),
the observed peak magnitude m$_{\rm max}$ (after correction for
line-of-sight reddening) is modified by a linear function of
the SN light curve decline rate $\Delta m_{15}$, where the latter
is a measure of the B-band decline (in magnitudes) from the peak of
the light curve to 15 days after the peak.  Higher-order quadratic
fits have also been employed on occasion (e.g., Suntzeff et~al. 1999;
Phillips et~al. 1999; Gibson et~al. 2000), as have more sophisticated
multicolour light curve shape techniques (Riess et~al. 1996,1998,1999).
All such modifications result in Hubble Diagram dispersions which
correspond to $\sim$7\% in distance, and are consistent with that expected 
from the photometric errors alone.

Starting with Mark Phillips' (2000) homogeneous re-analysis of the
photometry and light curves for both the
C-T and CfA SNe samples, we construct our desired B-, V-, and I-band
Hubble Diagrams by enforcing two strict criteria: the SNe must be
consistent with (i) 3.5$<$$\log(cz)_{\rm CMB}$$<$4.5, and (ii) $|$B$_{\rm
max}-$V$_{\rm max}$$|$$\le$0.20.  This yields a preferred subset of
36 SNe for the B- and V-band Hubble Diagrams, and 32 SNe for the I-band.
Criterion (i) is (essentially) universally adopted (for reasons already
noted previously), and (ii) was chosen
to mimic that employed by Hamuy et~al. (1996); stricter colour cuts have
also been used (e.g., Phillips et~al. 1999), but they have little impact
on the science which follows.

Fits in decline rate $\Delta m_{15}$ and peak magnitude $m_{\rm
max}$ are then constructed, the functional forms for which are
\begin{equation}
{\rm m}_{\rm max}-{\rm R}_{\rm m}{\rm E(B-V)}-5\log(cz) = 
c(\Delta m_{15}-1.1)^2+b(\Delta m_{15}-1.1)+a
\end{equation}
\noindent
where R$_{\rm m}$ is the assumed colour-dependent ratio of total-to-selective 
absorption, E(B-V) is the total line-of-sight reddening to a SN, and $cz$ is
its recessional velocity in the cosmic microwave background (CMB) reference 
frame.  The total line-of-sight reddening is decomposed into a 
Galactic foreground E(B-V)$_{\rm G}$ and host galaxy E(B-V)$_{\rm H}$ 
term:
\begin{equation}
{\rm E(B-V)} = {\rm E(B-V)_{\rm G}} + {\rm E(B-V)_{\rm H}}
\end{equation}
\noindent
where E(B-V)$_{\rm G}$ is provided by the COBE/DIRBE
dust maps of Schlegel et~al. (1998)
and E(B-V)$_{\rm H}$ is taken from Table~2 of Phillips et~al. (1999).
Prior to deriving the fit coefficients $a$, $b$, and $c$ for equation~(1),
extinction-dependent 
corrections to the decline rate (after Phillips et~al. 1999)
are applied, the form for which are
\begin{equation}
\Delta m_{15} = \Delta m_{15}({\rm raw}) + 
0.1[{\rm E(B-V)_{\rm G}} + {\rm E(B-V)_{\rm H}}].
\end{equation}

Figure~1 shows our culled sample of 36 SNe (32 in the I-band, recall) in the
peak magnitude-decline rate plane, after the extinction corrections of
equations~(2) and (3) have been applied.  A full accounting of the errors
in photometry, extinction, and CMB 
velocity - assumed to be $\pm$600\,km\,s$^{-1}$ -
(after Hamuy et~al. 1996, Phillips et~al. 1999, and Gibson et~al. 2000)
was made.

\begin{figure}[ht]
\plotfiddle{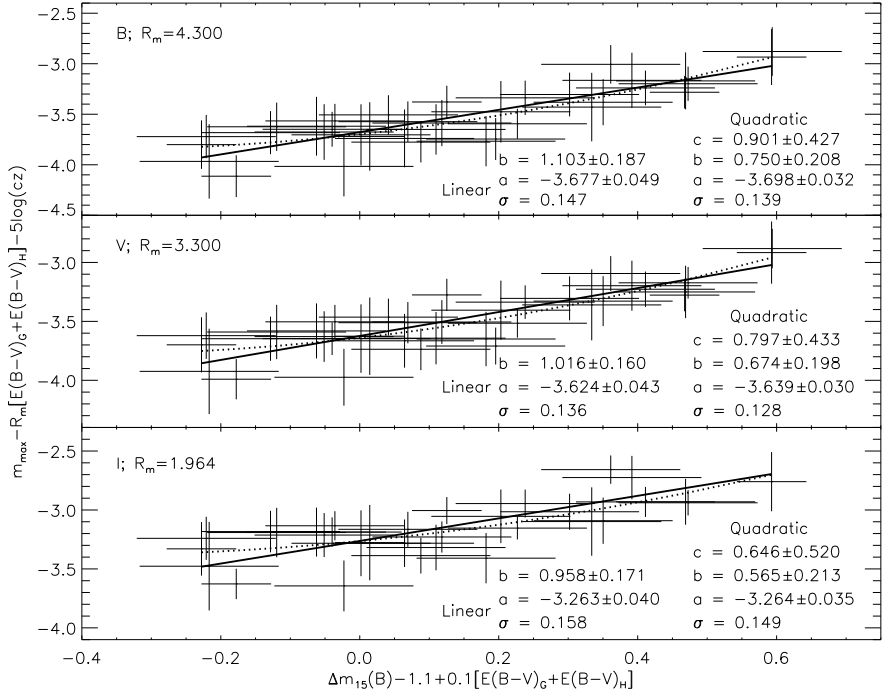}{3.35in}{0}{50}{50}{-165}{-10}
\caption{Peak magnitude m$_{\rm max}$ versus decline rate $\Delta m_{15}$
relationship for the subsample of C-T and CfA SNe with
3.5$<$$\log(cz)_{\rm CMB}$$<$4.5 and 
$|$B$_{\rm max}-$V$_{\rm max}$$|$$\le$0.20.  Foreground E(B-V)$_{\rm G}$
and host galaxy reddening corrections have been applied, after
Schlegel et~al. (1998) and Phillips et~al. (1999), respectively, assuming
ratios of total-to-selective absorption of R$_{\rm B}$:R$_{\rm V}$:R$_{\rm
I}$=4.300:3.300:1.964.  The velocities $cz$ are in the CMB reference frame.
Overlaid in each panel are the best quadratic (dotted) and linear (solid)
fits to the data; coefficients, their uncertainties, and the resulting
dispersion about the fit are likewise noted.
}
\end{figure}

A bootstrapped, quadratic, outlier-resistant,
fit to the data in each panel (B-, V-, and I-band,
from top to bottom) was then undertaken, employing the functional form
described by equation~(1).  The resultant colour-dependent coefficients
($a$, $b$, and $c$), their associated uncertainties, and overall dispersion 
$\sigma$ are shown under the heading ``Quadratic'' in Figure~1.  A bivariate
linear least squares fit was also employed (effectively forcing
coefficient $c$ of equation~(1) to be zero), the results for which
are also shown in Figure~1 adjacent to the heading ``Linear''.  While 
$\sigma_{\rm quad}$$<$$\sigma_{\rm lin}$, the statistical significance
of the improved fit is marginal (as measured by the Fisher F-test), and
is the primary reason why, for example, Freedman et~al. (2001) and Gibson \&
Stetson (2001) reverted to linear fits subsequent to our original
quadratic analysis (Gibson et~al. 2000).  Regardless, adopting the quadratic
fits in lieu of the linear ones has no impact upon the ultimately
derived value of the Hubble Constant.

A wide variety of fits were undertaken, above and beyond the canonical
quadratic and linear fits described above, including SNe subsets
comprised of C-T SNe alone, replacing the Schlegel et~al. (1998) foreground
extinction maps with the classical
Burstein \& Heiles (1982) values, reducing the recessional velocity uncertainty
from $\pm$600\,km\,s$^{-1}$ to (somewhat arbitrarily) $\pm$300\,km\,s$^{-1}$,
and modifying the 
ratios of total-to-selective absorption from\hfil\break
R$_{\rm B}$:R$_{\rm V}$:R$_{\rm
I}$=4.300:3.300:1.964 to R$_{\rm B}$:R$_{\rm V}$:R$_{\rm I}$=4.100:3.100:1.845
(the latter being the ratios employed by Hamuy et~al. 1996, Suntzeff et~al.
1999, and Phillips et~al. 1999).  All permutations led to (essentially)
identical values for the Hubble Constant, and so the reader will be spared
the painful details.

\section{Deriving the Hubble Constant}

Having generated what appears to be an excellent standard candle
(the corrected peak luminosity of a Type~Ia SNe), limited now
only by photometric
precision, we are left with the set of coefficients $a$, $b$, and (in the
case of quadratic fits) $c$ for equation~(1), but no immediate
value for the Hubble Constant.  The problem lies in the fact that
the corrected peak luminosity is indeed an excellent secondary distance
indicator (i.e., the corrected Hubble Diagrams provide accurate
\it relative \rm distances), but it still needs to be calibrated with
nearby SNe of \it known \rm distance $d$ (and therefore \it known \rm
luminosity M$_{\rm max}$).  Specifically, with equation~(1) and
some basic ASTRO~101:
\begin{eqnarray}
{\rm M}_{\rm max} & = & {\rm m}_{\rm max} - 5\log d - 25 \nonumber \\
cz_{\rm CMB}      & = & {\rm H}_0\,d \nonumber
\end{eqnarray}
\noindent
one is left with a set of colour-dependent quadratic (or linear)
equations relating the Hubble Constant H$_0$(B,V,I) to the peak
absolute magnitude M$_{\rm B,V,I}^{\rm max}$:
\begin{equation}
{\rm H}_0({\rm B,V,I}) = f({\rm M}_{\rm B,V,I}^{\rm max}).
\end{equation}
\noindent

Thanks primarily to the efforts of the
\it Sandage/Saha 
Type~Ia SN HST Calibration Project \rm (Saha et~al. 1999), accurate 
Cepheid-based distances can now be derived for ten nearby galaxies which 
play host to well-observed Type~Ia SNe.  The Sandage/Saha program includes
IC~4182 and NGCs~5253, 4536, 4496A, 4639, 3627, 
3982,\footnote{Data proprietary at the time of writing.} and 4527, and has
been supplemented with two additional calibrators
situated within NGC~3368 (Tanvir et~al. 1999) and NGC~4414 (Turner et~al.
1998).  Table~1 below lists the galaxy and SN
name for the seven highest-quality peak luminosity calibrators; it is these
seven which provide the M$_{\rm B,V,I}^{\rm max}$ 
calibration employed in determining H$_0$ via equation~(4).

The data reduction, Cepheid identification, and PL fitting for all but one of
the above galaxies (NGC~4527) has already been presented in Gibson et~al.
(2000), and so will not be replicated here.  Suffice it to say that the
data were processed with ALLFRAME (instrumental photometry - Stetson 1994) and
TRIAL (calibration and variable finding - Stetson 1996), following
the precepts laid out earlier by the
\it HST Key Project on the Extragalactic Distance Scale\rm.  The WFPC2
photometric zero point and charge-transfer corrections used here 
are based on Stetson (1998),\footnote{Which are nearly identical to
those of Whitmore et~al. (1999) and Dolphin (2000).} which differs from
the Gibson et~al. (2000) analysis where an \it 
a posteriori \rm systematic
transformation to the Hill et~al. (1998) zero point was applied.  
The justification for
adopting the Stetson (1998) calibration has been presented previously
(Freedman et~al. 2001; Gibson \& Stetson 2001).  

Two further modifications to the Gibson et~al. (2000) analysis are (i)
the adoption of an LMC true modulus of 18.45$\pm$0.10, and (ii)
the LMC apparent PL relations of Udalski et~al. (1999).  The justification
for both are provided by Freedman et~al. (2001); both (i) and (ii) result
in systematic downward shifts in derived galaxy distance, of 
0.05\,mag and (of order) 0.16\,mag, respectively.  The Udalski et~al. (1999)
I-band PL relation is $\sim$0.10\,mag\,dex$^{-1}$ flatter in slope than 
that claimed by Madore \& Freedman (1991), and is to be preferred over
the latter (Freedman et~al. 2001).  

Since the Gibson et~al. (2000) analysis, Gibson \& Stetson (2001) have
analysed NGC~4527, host galaxy to the spectroscopically peculiar SN~1991T.
Sixteen high-quality Cepheids were identified in the dataset, PL fitting
undertaken, and a true distance modulus of 30.482$\pm$0.085 derived
(neglecting any putative metallicity-dependent modifications to the 
Cepheid PL relations).  Figure~2 shows the V- and I-band PL relations for
the NGC~4527 Cepheids (upper two panels), and the distribution of individually 
de-reddened distance moduli (lower panel).

\begin{figure}[ht]
\plotfiddle{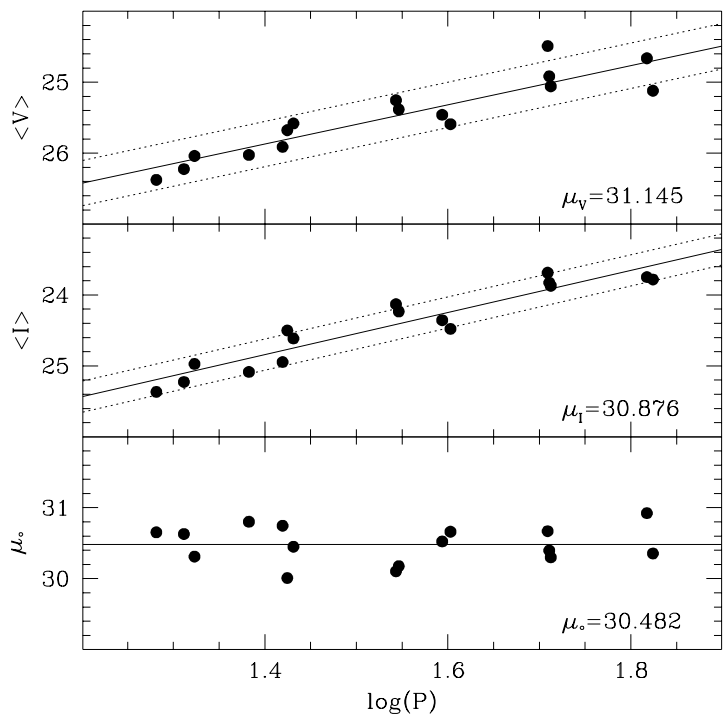}{3.35in}{0}{50}{50}{-150}{-85}
\caption{Apparent period-luminosity relations in the $V$-
(\it upper panel\rm) and $I$-bands (\it middle panel\rm) based upon the 16
high-quality Cepheid candidates discovered in NGC~4527 (Gibson \& Stetson 
2001).
The solid lines are least-squares fits to this entire sample, with the slope 
fixed to be that of the Udalski et~al. (1999) LMC
PL relations, while the dotted lines represent their corresponding 2$\sigma$
dispersion.  The inferred apparent distance moduli, ignoring metallicity
effects, are then $\mu_{\rm V}$=31.145$\pm$0.056 (internal) and $\mu_{\rm
I}$=30.876$\pm$0.046 (internal).
\it Lower Panel\rm: Distribution of individually de-reddened Cepheid
true moduli, as a function of period.  The mean corresponds
to $<\mu_\circ>$=30.482$\pm$0.066 (internal: $\pm$0.085\,mag,
incorporating all random uncertainties - after entry
R$_{\rm PL}$ in Table~7 of Gibson et~al. 2000).
}
\end{figure}

Table~1 summarises the results of our (Gibson \& Stetson 2001)
recent re-analysis of the extant
HST WFPC2 data for the host galaxies of the 
seven high-quality nearby Type~Ia SNe calibrators.  Re-fitting of Cepheid
PL relations with ($\mu_{\rm Z}$) and without ($\mu_\circ$) corrections
made for a putative metallicity-dependency to the PL relations - i.e.,
$\gamma_{\rm VI}$=$-$0.2\,mag\,dex$^{-1}$ and $\gamma_{\rm
VI}$=$+$0.0\,mag\,dex$^{-1}$, respectively - were undertaken.  The number of 
Cepheids employed in the fitting and the accompanying total random
uncertainties for each galaxy are also listed.

\begin{table}[ht]
\begin{center}
\caption{Distances to Nearby Type~Ia Supernovae Luminosity Calibrators and 
their Host Galaxies\\}
\begin{tabular}{llcccc}
\tableline
Galaxy & Supernova & n$_{\rm Ceph}$\tablenotemark{a} & 
$\mu_\circ$\tablenotemark{b} & 
$\mu_{\rm Z}$\tablenotemark{c} & $\sigma_\mu$(r)\tablenotemark{d} \\
\tableline
NGC~4527  & 1991T &      16 & 30.482 & 30.562 & $\pm$0.085 \\
NGC~4639  & 1990N &      17 & 31.524 & 31.624 & $\pm$0.084 \\
NGC~4536  & 1981B &      27 & 30.693 & 30.763 & $\pm$0.069 \\
NGC~3627  & 1989B &      17 & 29.794 & 29.944 & $\pm$0.169 \\
NGC~3368  & 1998bu& $\;\,$7 & 29.956 & 30.096 & $\pm$0.098 \\
NGC~5253  & 1972E & $\;\,$7 & 27.485 & 27.415 & $\pm$0.119 \\
IC~4182   & 1937C &      28 & 28.207 & 28.187 & $\pm$0.076 \\
\tableline
\tableline
\vspace{-11.0mm}
\tablenotetext{a}{Number of Cepheids employed in PL fitting.  The Cepheids
are identical to those used by Gibson et~al. (2000) and Gibson \& Stetson
(2001).}
\tablenotetext{b}{True distance moduli to Type~Ia SN host galaxies, assuming
Stetson (1998) WFPC photometric calibration, Udalski et~al. (1999) LMC PL
slopes and apparent zero points, and referenced to an LMC distance of
49\,kpc.}
\tablenotetext{c}{Metallicity-corrected distance moduli, assuming 
$\gamma_{\rm VI}$=$-$0.2\,mag\,dex$^{-1}$ and Cepheid field metallicities
as listed in Gibson et~al. (2000; Table~1) and Gibson \& Stetson (2001).}
\tablenotetext{d}{Random uncertainties are derived following Gibson
et~al. (2000), and correspond to item $R_{\rm PL}$ of Table~7 therein.}
\end{tabular}
\end{center}
\end{table}

Using the SN photometry and reddenings tabulated by Gibson et~al. (2000;
Table~5), supplemented now with the apparent peak magnitudes (Lira et~al.
1998; Table~7) and Galactic$+$intrinsic reddenings (Phillips et~al. 1999;
Table~2) for SN~1991T, the B-, V-, and I-band peak luminosities for
the seven calibrating SNe were then calculated.  While omitted for brevity's
sake, they are listed in columns
3-5 of Table~3 in Gibson \& Stetson (2001), assuming the metallicity-corrected 
true moduli $\mu_{\rm
Z}$ listed in Table~1.  It is important to note that the
corrected peak luminosity for
SN~1991T (e.g., M$_{\rm B,corr}^{\rm max}$=$-$19.40$\pm$0.24) is
indistinguishable from that of the mean of the full sample of nine
calibrators ($<$M$_{\rm B,corr}^{\rm max}$$>$=$-$19.32$\pm$0.08).  While
SN~1991T is 
only one datum, there exists no evidence to suggest that spectroscopically
peculiar Type~Ia SNe need be dismissed \it a priori \rm from future
extragalactic distance scale work.

The calibrated (corrected) peak luminosities can then
be used in conjunction with either 
the quadratic or linear fits presented in
Figure~1 (in conjunction with equations~1 and 4) to provide 
colour-dependent Hubble Constants for each of the seven calibrators.  
The weighted mean of H$_0$(B), H$_0$(V), and H$_0$(I) yields
H$_0$=73\,km\,s$^{-1}$\,Mpc$^{-1}$, with a total \it random \rm uncertainty of
$\pm$2\,km\,s$^{-1}$\,Mpc$^{-1}$ - both the quadratic and linear fits
of Figure~1 lead to the \it identical \rm result.\footnote{The formal
difference is at the $<$0.7\% level - i.e., $<$0.2$\sigma$!}

After Freedman et~al. (2001), seven sources of error were
incorporated into the \it systematic \rm error budget (and are
listed in Table~2).  Uncertainties in
the LMC zero point, crowding, and large scale bulk flows each enter in at
the $\pm$0.10\,mag level; the metallicity dependency of the Cepheid PL
relation at the $\pm$0.08\,mag level; the WFPC2 zero point uncertainty at
the $\pm$0.07\,mag level; reddening and bias in the Cepheid PL fitting at
the $\pm$0.02\,mag level each.  In quadrature, the overall systematic
error budget amounts to 0.21\,mag, corresponding to 10\% in H$_0$.
Significantly improving the precision to which we can derive H$_0$ via
Cepheid calibration of secondary distance indicators will require a factor
of two reduction in uncertainty in \it each \rm of these five remaining
dominant sources of systematic uncertainty.  Until then, we are limited to
10\% precision.

\begin{table}[ht]
\begin{center}
\caption{Systematic Error Budget\\}
\begin{tabular}{lc}
\tableline
Source of Uncertainty             & Error (mag) \\
\tableline
LMC true modulus                  & $\pm$0.10 \\
Stellar profile crowding on WFPC2 & $\pm$0.10 \\
Large-scale bulk flows            & $\pm$0.10 \\
Cepheid PL-[O/H] sensitivity      & $\pm$0.08 \\
WFPC2 calibration                 & $\pm$0.07 \\
Cepheid reddening                 & $\pm$0.02 \\
Bias in Cepheid PL relation       & $\pm$0.02 \\
{\it Total Systematic Uncertainty}& {\it $\pm$0.21} \\
\tableline
\tableline
\end{tabular}
\end{center}
\end{table}

In combination, the above random (r) and systematic (s) error budget yields
a final result for the Hubble Constant of
\begin{equation}
{\rm H}_0 = 73\pm 2({\rm r})\pm 7({\rm s})\;{\rm
km\,s^{-1}\,Mpc^{-1}}.
\end{equation}
\noindent
Ignoring the metallicity
dependency in the Cepheid PL relation (i.e., using $\gamma_{\rm
VI}$=$+$0.0\,mag\,dex$^{-1}$, as opposed to the $\gamma_{\rm
VI}$=$-$0.2$\pm$0.2\,mag\,dex$^{-1}$ employed here) increases H$_0$ by
3\,km\,s$^{-1}$\,Mpc$^{-1}$.

As emphasised earlier, a variety of fits to C-T and CfA SNe samples were
undertaken.  The canonical linear and quadratic fits in $\Delta m_{15}$
incorporated (i) the Schlegel et~al. (1998) and Phillips et~al. (1999)
prescriptions for foreground and host galaxy reddening, respectively; (ii)
extinction-dependent corrections to the decline rate, also after
Phillips et~al. (1999); (iii) recessional velocity uncertainties of
$\pm$600\,km\,s$^{-1}$; (iv) ratios of total-to-selective absorption of
R$_{\rm B}$:R$_{\rm V}$:R$_{\rm I}$=4.300:3.300:1.964.  In addition, the
combined dataset was culled from 51 to 36 (32 for the I-band), to avoid
highly-reddened and low-velocity (where the peculiar velocity uncertainty
becomes a substantial fraction of the recessional velocity itself) interlopers
biasing the analysis.

The colour-dependent coefficients 
provided by both the bivariate linear least squares and quadratic
bootstrapping fits in light curve decline rate
are robust against the adopted extinction law, extinction-dependent
corrections to the light curve shape, and host galaxy peculiar velocity
uncertainty.
The only two permutations which impacted equation~(5) at more than the 1\%
level were: (i) restricting the analysis to the 26 C-T SNe alone (as opposed to
the combined default sample of 36) reduces H$_0$ by $\sim$2\% (i.e.,
$\sim$1$\sigma$); (ii) completely neglecting the host galaxy reddening
correction (in both the C-T$+$CfA \it and \rm nearby calibrators) increases
H$_0$ by $\sim$4\% (i.e., $\sim$2$\sigma$).  The poor quality
of the linear and quadratic fits for case (ii) - e.g., the B-band
dispersion increases from $\sim$0.14\,mag to $\sim$0.21\,mag - 
makes the significance of this latter effect questionable at best.

\section{Summary}

A re-analysis of the combined Cal\'an-Tololo and Center for Astrophysics
Type~Ia SNe datasets is presented, with new linear and quadratic fits
to the empirical peak magnitude-decline rate relationship determined.
The fit coefficients are extremely robust to uncertainties in photometry, 
reddening, and peculiar velocity.  With a calibration to the corrected
Hubble Diagrams now provided by seven high-quality nearby calibrators with
accurate host galaxy Cepheid distance determinations, our favoured value for
the Hubble Constant becomes:
\begin{center}
\fbox{${\rm H}_0 = 73\pm 2({\rm r})\pm 7({\rm s})\;{\rm km\,s^{-1}\,Mpc^{-1}}$}
\end{center}
\noindent
The systematic uncertainties clearly limit the precision to which we can
determine H$_0$; until significant progress is made in improving our
understanding of, for example, the distance to the LMC and the sensitivity of
the Cepheid PL relation to metallicity, the uncertainty associated with
Cepheid-based Hubble Constant \it cannot \rm be reduced below 10\%.

\acknowledgements
We wish to thank Peter Stetson for his crucial contributions throughout the
course of this work.


\end{document}